\documentclass[10pt,a4paper]{article}
\usepackage{amsmath}
\usepackage{float}
\usepackage{caption2}
\usepackage{graphicx}
\def\openone{\leavevmode\hbox{\normalsize1\kern-3.6pt\normalsize1}}

\begin{document}

\title{\textbf{\Large{On the equivalence of the CH and CHSH inequalities for two\\
\vspace{-.2cm}three-level systems\thanks{This paper has been originally published in: International Journal of Quantum Information \textbf{1}, 115-133 (2003).}}}}
\author{Jos\'{e} L.\ Cereceda\thanks{Electronic mail: jl.cereceda@teleline.es} \\
\textit{C/Alto del Le\'{o}n 8, 4A, 28038 Madrid, Spain}}

\date{January 6, 2004}

\maketitle

\begin{abstract}
In this paper we show a Clauser-Horne (CH) inequality for two three-level quantum systems or qutrits, alternative to the CH inequality given by Kaszlikowski \textit{et al.} [Phys. Rev. A \textbf{65}, 032118 (2002)]. In contrast to this latter CH inequality, the new one is shown to be equivalent to the Clauser-Horne-Shimony-Holt (CHSH) inequality for two qutrits given by Collins \textit{et al.} [Phys. Rev. Lett. \textbf{88}, 040404 (2002)]. Both the CH and CHSH inequalities exhibit the strongest resistance to noise for a nonmaximally entangled state for the case of two von Neumann measurements per site, as first shown by Ac\'\i n \textit{et al.} [Phys. Rev. A \textbf{65}, 052325 (2002)]. This equivalence, however, breaks down when one takes into account the less-than-perfect quantum efficiency of detectors. Indeed, for the noiseless case, the threshold quantum efficiency above which there is no local and realistic description of the experiment for the optimal choice of measurements is found to be $(9-\sqrt{33})/4 \approx 0.814$ for the CH inequality, whereas it is equal to $(\sqrt{-3+\sqrt{33}})/2 \approx 0.828$ for the CHSH inequality.

\vspace{.2cm}

\end{abstract}

\section{Introduction and notation}

Recently, two kinds of Bell inequalities \cite{Bell64} have been introduced for two three-dimensional quantum systems (so-called qutrits). The scenario for both types of inequalities involves two parties: Alice can carry out two possible measurements, $A_1$ or $A_2$, on one of the qutrits, whereas Bob is allowed to perform the measurements $B_1$ or $B_2$ on the other qutrit. Each measurement has three possible outcomes $A_i, B_j = 1,2,3$ ($i,j=1,2$). Then, denoting by $P(A_i = B_j +k)$ the probability that the measurements $A_i$ and $B_j$ have outcomes that differ by $k$ modulo $d$ (in our case $d=3$), the Collins \textit{et al.} \cite{CGLMP} arrived at the following Bell inequality:
\begin{align}
I_3 = \,\, & P(A_1=B_1)+P(B_1=A_2+1)+P(A_2=B_2)+P(B_2=A_1)  \nonumber  \\
& -P(A_1=B_1-1)-P(B_1=A_2)-P(A_2=B_2-1)  \nonumber  \\
& -P(B_2=A_1-1) \leq 2 .
\label{I3}
\end{align}
The Bell inequality (1) is of the CHSH type because it reduces to the familiar CHSH inequality \cite{CHSH} for $d=2$. Actually, inequality (\ref{I3}) is a particular case of the family of Bell inequalities (CGLMP-set):
\begin{align}
I_3(c_1,c_2,c_3,c_4) &= P(A_1 =B_1+c_1) +P(B_1=A_2+c_2) +P(A_2=B_2+c_3) \nonumber  \\
 +\, P(B_2 &=A_1+c_4) -P(A_1=B_1-(c_2+c_3+c_4)) -P(B_1=A_2-(c_1+c_3+c_4))\nonumber  \\
 -\, P(A_2 &=B_2-(c_1+c_2+c_4)) -P(B_2=A_1-(c_1+c_2+c_3)) \leq 2,
\label{CGLMPset}
\end{align}
where
\begin{gather}
c_i=0,\pm 1,  \nonumber  \\
c_1+c_2+c_3+c_4 \neq 0\,\text{mod}3,  \nonumber
\end{gather}
and where the sum is modulo 3 for the $c_i$'s. Inequality (\ref{I3}) is obtained for $c_2=+1$ and $c_1=c_3=c_4=0$. There are 54 combinations of $c_i$'s (with $c_i=0,\pm 1$) fulfilling the condition $c_1+c_2+c_3+c_4 \neq 0\,\text{mod}3$.

On the other hand, denoting by $P^{ij}(a_i,b_j)$ the joint probability of obtaining by Alice and Bob simultaneously the results $a_i$ and $b_j$ ($a_i,b_j=1,2,3$) for the pair of observables $A_i$ and $B_j$, and denoting by $P^{i}(a_i)$ ($Q^{j}(b_j)$) the single probability of obtaining the result $a_i$ ($b_j$) by Alice (Bob) irrespective of Bob's (Alice's) outcome, the Kaszlikowski \textit{et al.} \cite{KKCZO} arrived at the following Bell inequality:\footnote{
\small{The Bell inequality (\ref{W3}) is a member of the set of CH inequalities introduced in \cite{KKCZO}, namely,
\begin{align}
& P^{1+\alpha\, 1+\beta}(2+x,1+y)+P^{1+\alpha\, 2+\beta}(2+x,1+y)
  -P^{2+\alpha\, 1+\beta}(2+x,1+y) \nonumber \\
& +P^{2+\alpha\, 2+\beta}(2+x,1+y)+ P^{1+\alpha\, 1+\beta}(1+x,2+y)
  +P^{1+\alpha\, 2+\beta}(1+x,2+y)  \nonumber \\
& -P^{2+\alpha\, 1+\beta}(1+x,2+y)+P^{2+\alpha\, 2+\beta}(1+x,2+y) 
+ P^{1+\alpha\, 1+\beta}(2+x,2+y)  \nonumber  \\
& +P^{1+\alpha\, 2+\beta}(1+x,1+y)-P^{2+\alpha\, 1+\beta}(2+x,2+y)
+ P^{2+\alpha\, 2+\beta}(2+x,2+y)  \nonumber  \\  
& - P^{1+\alpha}(1+x)-P^{1+\alpha}(2+x)-Q^{2+\beta}(1+y)-Q^{2+\beta}(2+y)  \leq 0,  \nonumber
\end{align}
where $\alpha,\beta=0,1$; $x,y=0,1,2$, and where the addition is modulo 2 for $\alpha,\beta$ and modulo 3 for $x,y$. Inequality (\ref{W3}) is obtained for $\alpha,\beta=0$ and $x,y=0$.}}
\begin{align}
W_3  =\,\, & P^{11}(2,1)+P^{12}(2,1)-P^{21}(2,1)+P^{22}(2,1)  \nonumber  \\
& + P^{11}(1,2)+P^{12}(1,2)-P^{21}(1,2)+P^{22}(1,2)  \nonumber  \\
& + P^{11}(2,2)+P^{12}(1,1)-P^{21}(2,2)+P^{22}(2,2)  \nonumber  \\  
& - P^{1}(1)-P^{1}(2)-Q^{2}(1)-Q^{2}(2)  \leq 0.
\label{W3}
\end{align}
As noticed in Ref.~\cite{KKCZO}, the Bell inequality (\ref{W3}) is the sum of two CH inequalities plus one term which bears a resemblance to an incomplete CH inequality. (Previous derivations of Bell inequalities for two three-level systems based on the original Clauser-Horne inequalities \cite{CH} can be found, for example, in Refs.~\cite{ZZH} and \cite{WZPW}.) It has been shown that both the CHSH inequality (\ref{I3}) and the CH inequality (\ref{W3}) give the same threshold value of noise admixture (for which it is still not possible to build a local classical model for the predicted probabilities) for the maximally entangled state \cite{CGLMP,KKCZO} (see also Refs.\ \cite{KGZMZ,DKZ,CKKZO}). This notwithstanding, as we will show, inequalities (\ref{I3}) and (\ref{W3}) are \textit{not\/} equivalent. This might seem rather surprising in view of the fact that, for bipartite two-dimensional systems, the familiar CHSH and CH inequalities are equivalent provided that the correlations cannot be used for instantaneous communication between Alice and Bob \cite{Mermin,Cereceda}. So a non-trivial question is whether the set of CH inequalities introduced in Ref.\ \cite{KKCZO} does exhaust all possible instances of CH inequalities for two three-level systems. In this paper we answer this question---a negative one---by exhibiting a CH inequality having the same structure as inequality (\ref{W3}), and which is equivalent to the CHSH inequality (\ref{I3}). Indeed, it will be argued that to each of the CHSH inequalities in the CGLMP-set (\ref{CGLMPset}), there corresponds one and only one  independent CH inequality which is equivalent to it.

Central to the derivation of our results is the above-mentioned property of causal communication (also termed ``physical locality'' \cite{Mermin}). For the experiment considered, this means that the marginal probabilities for one party should be independent of the measurement chosen by the other party:
\begin{equation}
\sum_{n=1}^{3} P(A_i=m,B_1=n) = \sum_{n=1}^{3} P(A_i=m,B_2=n),
\label{NSC1}
\end{equation}
and
\begin{equation}
\sum_{m=1}^{3} P(A_1=m,B_j=n) = \sum_{m=1}^{3} P(A_2=m,B_j=n),
\label{NSC2}
\end{equation}
for any $i,j=1,2$ and $m,n=1,2,3$. The fulfillment of (\ref{NSC1}) and (\ref{NSC2}) constitutes a physically sound requirement since a violation of either (\ref{NSC1}) or (\ref{NSC2}) would, in principle, allow the two parties to communicate superluminally. Both quantum mechanics and classical theories satisfy the requirement of causal communication (the ``no-signaling'' condition), and hence the predictions by such theories do satisfy each of the constraints in (\ref{NSC1}) and (\ref{NSC2}). In addition to this, the joint probabilities are required to satisfy the normalization condition:
\begin{equation}
\sum_{m,n=1}^{3} P(A_i=m,B_j=n) = 1,
\label{NC}
\end{equation}
for any $i,j=1,2$.

The paper is organized as follows. In Sec.\ 2, we provide a CH-type inequality for bipartite systems of qutrits, alternative to the original CH inequality introduced in Ref.~\cite{KKCZO}. Using the conditions in (\ref{NSC1})--(\ref{NC}), we show that the given CH inequality is equivalent to the CHSH inequality (\ref{I3}). We point out that, actually, a similar relationship could be established between each CHSH inequality in the set (\ref{CGLMPset}) and some appropriate CH inequality. Conditions (\ref{NSC1})--(\ref{NC}) are also used to show that inequalities (\ref{I3}) and (\ref{W3}) are not equivalent. In Sec.\ 3, we describe the optimal set of measurements giving the maximal violation of both the CH and CHSH inequalities, and determine the resistance to noise of such inequalities for the optimal choice of observables. In Sec.\ 4, we consider the realistic case of detectors with a finite quantum detector efficiency $\eta <1$. As we will see, for this case the equivalence between the CH and CHSH inequalities does not follow any more. For the noiseless case, we calculate the critical quantum efficiency needed to rule out a local and realistic description of the considered experiment for both the CH and CHSH inequalities. Finally, the main conclusions are summarized in Sec.~5.

In order to abbreviate the notation, we will henceforth use at our convenience the following shorthand notation for the various joint probabilities:
\begin{align}
\qquad p_1& \equiv P^{11}(1,1),  & p_2& \equiv P^{11}(1,2),  &  p_3& \equiv P^{11}(1,3),
 &  p_4& \equiv P^{11}(2,1), \qquad \nonumber \\
\qquad p_5& \equiv P^{11}(2,2),  &   p_6& \equiv P^{11}(2,3),  & p_7& \equiv P^{11}(3,1),
 & p_8& \equiv P^{11}(3,2), \qquad \nonumber \\
\qquad p_9& \equiv P^{11}(3,3),  & p_{10}& \equiv P^{12}(1,1),  &  p_{11}& \equiv P^{12}(1,2),
 &  p_{12}& \equiv P^{12}(1,3), \qquad \nonumber \\
\qquad p_{13}& \equiv P^{12}(2,1),  &   p_{14}& \equiv P^{12}(2,2),  & p_{15}& \equiv P^{12}(2,3),
 & p_{16}& \equiv P^{12}(3,1), \qquad \nonumber \\
\qquad p_{17}& \equiv P^{12}(3,2),  & p_{18}& \equiv P^{12}(3,3),  &  p_{19}& \equiv P^{21}(1,1),
 &  p_{20}& \equiv P^{21}(1,2), \qquad  \label{notation}  \\
\qquad p_{21}& \equiv P^{21}(1,3),  &  p_{22}& \equiv P^{21}(2,1),  & p_{23}& \equiv P^{21}(2,2),
 & p_{24}& \equiv P^{21}(2,3), \qquad \nonumber \\
\qquad p_{25}& \equiv P^{21}(3,1),  & p_{26}& \equiv P^{21}(3,2),  &  p_{27}& \equiv P^{21}(3,3),
 &  p_{28}& \equiv P^{22}(1,1), \qquad \nonumber \\
\qquad p_{29}& \equiv P^{22}(1,2),  &   p_{30}& \equiv P^{22}(1,3),  & p_{31}& \equiv P^{22}(2,1),
 & p_{32}& \equiv P^{22}(2,2), \qquad \nonumber \\
\qquad p_{33}& \equiv P^{22}(2,3),  &   p_{34}& \equiv P^{22}(3,1),  & p_{35}& \equiv P^{22}(3,2),
 & p_{36}& \equiv P^{22}(3,3). \qquad \nonumber
\end{align}

\section{Alternative CH inequality}

Let us consider the inequality
\begin{align}
K_3  = \,\, & P^{11}(1,1)+P^{12}(1,1)-P^{21}(1,1)+P^{22}(1,1)  \nonumber  \\
& + P^{11}(2,2)+P^{12}(2,2)-P^{21}(2,2)+P^{22}(2,2)  \nonumber  \\
& + P^{11}(2,1)+P^{12}(1,2)-P^{21}(2,1)+P^{22}(2,1)  \nonumber  \\  
& - P^{1}(1)-P^{1}(2)-Q^{2}(1)-Q^{2}(2)  \leq 0.
\label{K3}
\end{align}
As in the case of the CH inequality (\ref{W3}), the inequality (\ref{K3}) can be written as the sum of two CH inequalities, $\text{CH}_1$ and $\text{CH}_2$, and some additional term, with $\text{CH}_1$ being
\begin{equation*}
P^{11}(1,1)+P^{12}(1,1)-P^{21}(1,1)+P^{22}(1,1)-P^{1}(1)-Q^{2}(1),
\end{equation*}
and $\text{CH}_2$ being
\begin{equation*}
P^{11}(2,2)+P^{12}(2,2)-P^{21}(2,2)+P^{22}(2,2)-P^{1}(2)-Q^{2}(2).
\end{equation*}
Note that the single probabilities appearing in (\ref{K3}) are the same as those appearing in (\ref{W3}).

In what follows we show that the CH inequality $K_3 \leq 0$ is equivalent to the CHSH inequality $I_3 \leq 2$ in (\ref{I3}). More precisely, using relations (\ref{NSC1})--(\ref{NC}), we show that the left hand side of inequality (\ref{K3}) can be expressed as $K_3= (I_3-2)/3$. Clearly, the inequality $I_3 \leq 2$, then implies the inequality $K_3 \leq 0$. Conversely, starting from the left hand side of inequality (\ref{I3}), and using relations (\ref{NSC1})--(\ref{NC}), we show that $I_3 =2+3K_3$. Therefore, it is also the case that the inequality $K_3 \leq 0$ implies the inequality $I_3 \leq 2$. To this end, we first write both $I_3$ and $K_3$ as a sum of joint probabilities. From (\ref{I3}), it can readily be seen that
\begin{align}
I_3 = \,\, & p_1+p_5+p_9+p_{10}+p_{14}+p_{18}+p_{20}
+p_{24}+p_{25}+p_{28}+p_{32}+p_{36}  \nonumber  \\
& -\big( p_2+p_6+p_7+p_{12}+p_{13}+p_{17}+p_{19}
+p_{23}+p_{27}+p_{29}+p_{33}+p_{34} \big) ,
\label{I3p}
\end{align}
where we have used the notation in (\ref{notation}). On the other hand, putting the single probabilities $P^{1}(1)$, $P^{1}(2)$, $Q^{2}(1)$, and $Q^{2}(2)$ as\footnote{
\small{We note that, since the joint probabilities $P^{ij}(a_i,b_j)$ satisfy the conditions in Eqs.\ (\ref{NSC1})--(\ref{NSC2}), the single probabilities $P^{i}(a_i)$ and $Q^{j}(b_j)$ can actually be expressed in terms of $P^{ij}(a_i,b_j)$ in two equivalent forms. So, for example, $P^{1}(1)$ in (\ref{link1}) can be put alternatively as, 
$P^{1}(1) = P^{11}(1,1)+P^{11}(1,2)+P^{11}(1,3)$, $Q^{2}(1)$ as $Q^{2}(1) = P^{12}(1,1)+P^{12}(2,1)
+P^{12}(3,1)$, etc. Of course, the result that $K_3= (I_3-2)/3$ or $I_3 =2+3K_3$ can be obtained by using either one of the two equivalent forms for each $P^{i}(a_i)$ and $Q^{j}(b_j)$. The advantage of using the choice in (\ref{link1}) is that it leads to an expression for $K_3$ involving only ten joint probabilities.}}
\begin{align}
P^{1}(1) = \,\, & P^{12}(1,1)+P^{12}(1,2)+P^{12}(1,3),  \nonumber  \\
P^{1}(2) = \,\, & P^{11}(2,1)+P^{11}(2,2)+P^{11}(2,3),  \nonumber  \\[-.3cm]
&  \label{link1} \\[-.3cm]
Q^{2}(1) = \,\, & P^{22}(1,1)+P^{22}(2,1)+P^{22}(3,1),  \nonumber  \\
Q^{2}(2) = \,\, & P^{22}(1,2)+P^{22}(2,2)+P^{22}(3,2),  \nonumber
\end{align}
and susbstituting (\ref{link1}) into the left hand side of (\ref{K3}), we obtain
\begin{equation}
K_3 = p_1-p_6-p_{12}+p_{14}-p_{19}-p_{22}-p_{23}-p_{29}-p_{34}-p_{35} .
\label{K3p}
\end{equation}
Of course, due to the constraints in (\ref{NSC1})--(\ref{NC}), not all the probabilities $p_1,p_2,\ldots,p_{36}$ are independent.  Now, in order to compare expressions (\ref{I3p}) and (\ref{K3p}), we have to know the relations that can be established between such probabilities. The normalization (\ref{NC}) plus no-signaling conditions (\ref{NSC1})--(\ref{NSC2}) constitute a linear system of 16 equations and 36 unknowns $p_1,p_2,\ldots,p_{36}$. It can be shown \cite{Masanes} that such a system determines 12 probabilities at most among $p_1,p_2,\ldots,p_{36}$. So, for example, we can solve the system of equations (\ref{NSC1})--(\ref{NC}) with respect to the set of variables $\{p_3,p_4,p_8,p_{11},p_{15},p_{16},p_{21},p_{22},p_{26},p_{30},p_{31},p_{35}\}$ to find, in particular, that
\begin{align}
p_{22} = \,\, & \frac{1}{3} \big( 1+p_1-p_2-2p_5-p_6+2p_7+p_9+p_{10}-p_{12}
      +2p_{13}+p_{14}-p_{17}-2p_{18} \nonumber \\ 
    & -p_{19}+p_{20}-p_{23}-2p_{24}-2p_{25}-p_{27}
      -2p_{28}-p_{29}+p_{32}+2p_{33}-p_{34}+p_{36} \big), \label{p22}
\end{align}
and
\begin{align}
p_{35} = \,\, & \frac{1}{3} \big( 1+p_1+2p_2+p_5-p_6-p_7-2p_9-2p_{10}-p_{12}
      -p_{13}+p_{14}+2p_{17}+p_{18} \nonumber \\ 
    & -p_{19}-2p_{20}-p_{23}+p_{24}+p_{25}+2p_{27}
      +p_{28}-p_{29}-2p_{32}-p_{33}-p_{34}-2p_{36} \big). \label{p35} 
\end{align}
Thus, inserting (\ref{p22}) and (\ref{p35}) into (\ref{K3p}) gives $K_3=(I_3 -2)/3$, with $I_3$ being the expression in (\ref{I3p}).

Alternatively, we can solve the system of equations (\ref{NSC1})--(\ref{NC}) with respect to the set of variables $\{p_2,p_4,p_9,\linebreak p_{11},p_{13},p_{18},p_{20},p_{24},p_{25},p_{28},p_{32},p_{36}\}$. In the Appendix we write down the resulting expressions for the probabilities $p_2$, $p_9$, $p_{13}$, $p_{18}$, $p_{20}$, $p_{24}$, $p_{25}$, $p_{28}$, $p_{32}$, and $p_{36}$ (see (A.1)--(A.10) in the Appendix). Substituting now (A.1)--(A.10) into (\ref{I3p}) we obtain $I_3=2+3K_3$, with $K_3$ being the expression in (\ref{K3p}).

Summing up, from the conditions of normalization and causal communication we have derived the relations $K_3=(I_3 -2)/3$ and $I_3=2+3K_3$, hence it follows the equivalence of the Bell inequalities: $K_3 \leq 0$ and $I_3 \leq 2$. We would like to emphasize that the no-signaling condition is satisfied by both quantum mechanics and local realistic theories, so the relation $K_3=(I_3 -2)/3$ (or $I_3=2+3K_3$) is certainly fulfilled by the probabilities predicted by such theories.

In the same way, one could equally show that each of the CHSH inequalities in the CGLMP-set (\ref{CGLMPset}) is associated with some appropriate CH inequality. So, for example, consider the inequality in (\ref{CGLMPset}) for which $c_4=+1$ and $c_1=c_2=c_3=0$:
\begin{align}
I_3^{\prime} = \,\, & P(A_1=B_1)+P(B_1=A_2)+P(A_2=B_2)+P(B_2=A_1+1)  \nonumber  \\
& -P(A_1=B_1-1)-P(B_1=A_2-1)-P(A_2=B_2-1)  \nonumber  \\
& -P(B_2=A_1) \leq 2 .
\label{I3prime}
\end{align}
In terms of joint probabilities, $I_3^{\prime}$ can be written in the form
\begin{align}
I_3^{\prime} = \,\, & p_1+p_5+p_9+p_{11}+p_{15}+p_{16}+p_{19}
+p_{23}+p_{27}+p_{28}+p_{32}+p_{36}  \nonumber  \\
& -\big( p_2+p_6+p_7+p_{10}+p_{14}+p_{18}+p_{21}
+p_{22}+p_{26}+p_{29}+p_{33}+p_{34} \big) .
\label{I3primep}
\end{align}
Consider now the CH-type inequality
\begin{align}
K_3^{\prime}  = \,\, & P^{11}(1,1)-P^{12}(1,1)+P^{21}(1,1)+P^{22}(1,1)  \nonumber  \\
& + P^{11}(2,2)-P^{12}(2,2)+P^{21}(2,2)+P^{22}(2,2)  \nonumber  \\
& + P^{11}(2,1)-P^{12}(2,1)+P^{21}(1,2)+P^{22}(2,1)  \nonumber  \\  
& - P^{2}(1)-P^{2}(2)-Q^{1}(1)-Q^{1}(2)  \leq 0.
\label{K3prime}
\end{align}
We mention, incidentally, that inequality (\ref{K3prime}) contains the following two CH inequalities: 
\begin{equation*}
\text{CH}_1 = P^{11}(1,1)-P^{12}(1,1)+P^{21}(1,1)+P^{22}(1,1)- P^{2}(1)-Q^{1}(1),
\end{equation*}
and 
\begin{equation*}
\text{CH}_2 = P^{11}(2,2)-P^{12}(2,2)+P^{21}(2,2)+P^{22}(2,2)- P^{2}(2)-Q^{1}(2).
\end{equation*}
Putting the single probabilities $P^{2}(1)$, $P^{2}(2)$, $Q^{1}(1)$, and $Q^{1}(2)$ as
\begin{align}
P^{2}(1) = \,\, & P^{21}(1,1)+P^{21}(1,2)+P^{21}(1,3),  \nonumber  \\
P^{2}(2) = \,\, & P^{22}(2,1)+P^{22}(2,2)+P^{22}(2,3),  \nonumber  \\[-.3cm]
&  \label{link2} \\[-.3cm]
Q^{1}(1) = \,\, & P^{11}(1,1)+P^{11}(2,1)+P^{11}(3,1),  \nonumber  \\
Q^{1}(2) = \,\, & P^{21}(1,2)+P^{21}(2,2)+P^{21}(3,2),  \nonumber
\end{align}
and substituting (\ref{link2}) into the left hand side of (\ref{K3prime}), we obtain
\begin{equation}
K_3^{\prime} = p_5-p_7-p_{10}-p_{13}-p_{14}-p_{20}-p_{21}-p_{26}+p_{28}-p_{33} .
\label{K3primep}
\end{equation}
In order to relate the expression for $K_3^{\prime}$ in (\ref{K3primep}) with the quantity $I_3^{\prime}$ in (\ref{I3primep}), we solve the system of equations (\ref{NSC1})--(\ref{NC}) with respect to the set of variables $\{p_3,p_4,p_8,p_{12},p_{13},p_{17},p_{20},p_{24},p_{25},p_{30},p_{31},p_{35}\}$. This gives, in particular,
\begin{align}
p_{13} = \,\, & \frac{1}{3} \big( 1-2p_1-p_2+p_5+2p_6-p_7+p_9-p_{10}+p_{11}
      -p_{14}-2p_{15}-2p_{16}-p_{18} \nonumber \\ 
    & +p_{19}-p_{21}+2p_{22}+p_{23}-p_{26}-2p_{27}
      +p_{28}-p_{29}-2p_{32}-p_{33}+2p_{34}+p_{36} \big), \label{p13}
\end{align}
and
\begin{align}
p_{20} = \,\, & \frac{1}{3} \big( 1+p_1+2p_2+p_5-p_6-p_7-2p_9-p_{10}-2p_{11}
      -p_{14}+p_{15}+p_{16}+2p_{18} \nonumber \\ 
    & -2p_{19}-p_{21}-p_{22}-2p_{23}-p_{26}+p_{27}
      +p_{28}+2p_{29}+p_{32}-p_{33}-p_{34}-2p_{36} \big). \label{p20}
\end{align}
Thus, using (\ref{p13}) and (\ref{p20}), in (\ref{K3primep}), we get $K_3^{\prime} = (I_3^{\prime}-2)/3$. On the other hand, solving with respect to the variables $\{p_1,p_6,p_8,p_{11},p_{15},p_{16},p_{19},p_{23},p_{27},p_{30},p_{32},p_{34}\}$, and replacing the resulting probabilities $p_1$, $p_6$, $p_{11}$, $p_{15}$, $p_{16}$, $p_{19}$, $p_{23}$, $p_{27}$, $p_{32}$, and $p_{34}$ in (\ref{I3primep}), we would find that $I_3^{\prime} =2+3K_3^{\prime}$, with $K_3^{\prime}$ being the expression in (\ref{K3primep}). Therefore, the inequality $I_3^{\prime}\leq 2$ implies the inequality $K_3^{\prime}\leq 0$, and conversely, the inequality $K_3^{\prime}\leq 0$ implies the inequality $I_3^{\prime}\leq 2$.

It is important to note that a given CHSH inequality can only be related to one independent CH inequality. To see this, suppose instead that the CHSH inequality $I_3\leq 2$ is related to two independent CH inequalities, $K_3\leq 0$ and $K_3^{\prime}\leq 0$, through the respective relations $I_3 =2+3K_3$ and $I_3=2+3K_3^{\prime}$. Then it trivially follows from such relations that $K_3 =K_3^{\prime}$, and hence the initial supposition that $K_3$ and $K_3^{\prime}$ are independent cannot be true. Analogously, it follows that a given CH inequality can only be related to one independent CHSH inequality. It should be noticed, however, that there is not a one-to-one correspondence between the set of CHSH inequalities in (\ref{CGLMPset}) and the set of CH inequalities since, for example, as we argue in the next paragraph, the CH inequality (\ref{W3}) is not equivalent to any of the inequalities in (\ref{CGLMPset}).

We now show that the CH inequality (\ref{W3}) and the CHSH inequality (\ref{I3}) are not equivalent. To see this, we first write $W_3$ in the equivalent form
\begin{equation}
W_3 = p_2-p_6-p_{10}-p_{12}-p_{16}-p_{20}-p_{22}-p_{23}+p_{31}-p_{35} .
\label{W3p}
\end{equation}
Then, solving the system of equations (\ref{NSC1})--(\ref{NC}) with respect to the set of variables $\{p_3,p_4,p_8,p_{11},p_{15},p_{16},p_{21},\linebreak p_{22},p_{26},p_{30},p_{31},p_{35}\}$, and substituting the resulting probabilities $p_{16}$, $p_{22}$, $p_{31}$, and $p_{35}$ into (\ref{W3p}), we would obtain
\begin{equation}
W_3 = \frac{1}{3} \big( I_3 -2 \big) -p_1 +p_2 +p_{13}-p_{14}+p_{19}-p_{20}-p_{28}+p_{29}.
\label{W3no}
\end{equation}
On the other hand, solving with respect to the variables
\begin{equation*}
\{p_1,p_5,p_9,p_{11},p_{13},p_{18},p_{19},p_{24},p_{26},p_{28},p_{32},p_{36}\},
\end{equation*}
and replacing the resulting probabilities $p_1$, $p_5$, $p_{9}$, $p_{13}$, $p_{18}$, $p_{19}$, $p_{24}$, $p_{28}$, $p_{32}$, and $p_{36}$ in (\ref{I3p}), we would obtain
\begin{equation}
I_3 = 2 +3 \big(-p_4-p_6-p_7-p_{12}+p_{14}-p_{23}+p_{25}-p_{29}-p_{34}-p_{35} \big),
\label{I3no1}
\end{equation}
which can be written equivalently as
\begin{equation}
I_3 = 2 +3W_3 +3\big(p_1 -p_2 -p_{13}+p_{14}-p_{19}+p_{20}+p_{28}-p_{29}\big) ,
\label{I3no2}
\end{equation}
with $W_3$ being the expression in (\ref{W3p}), and where we have made use of the relations
\begin{equation*}
p_{25}-p_4-p_7 = p_1 -p_{19}-p_{22},
\end{equation*}
and
\begin{equation*}
p_{10}+p_{16}-p_{31}-p_{34}=p_{28}-p_{13},
\end{equation*}
[cf.~(\ref{NSC2})] in passing from (\ref{I3no1}) to (\ref{I3no2}). Of course, the expression for $I_3$ in (\ref{I3no2}) can also be obtained directly from (\ref{W3no}). From this latter equation we can deduce that the inequality $I_3\leq 2$ would imply the inequality $W_3 \leq0$ provided that, for $I_3\leq 2$, the following inequality:
\begin{equation}
p_2+p_{13}+p_{19}+p_{29} \leq p_1+p_{14}+p_{20}+p_{28},
\label{C1no}
\end{equation}
is satisfied. On the other hand, from (\ref{I3no2}), it can be seen that the inequality $W_3 \leq 0$ would imply the inequality $I_3 \leq 2$ provided that, for $W_3\leq 0$, the following inequality:
\begin{equation}
p_1+p_{14}+p_{20}+p_{28} \leq p_2+p_{13}+p_{19}+p_{29},
\label{C2no}
\end{equation}
is satisfied. Obviously, the conditions in (\ref{C1no}) and (\ref{C2no}) are mutually exclusive except for the particular event in which
\begin{equation*}
p_1+p_{14}+p_{20}+p_{28} = p_2+p_{13}+p_{19}+p_{29}.
\end{equation*}
This means that it is in general not possible for the inequality $W_3 \leq 0$ to imply the inequality $I_3 \leq 2$, and simultaneously for the inequality $I_3 \leq 2$ to imply the inequality $W_3 \leq 0$. In other words, such inequalities are not equivalent. Similarly, it could equally be shown that the CH inequality (\ref{W3}) is not equivalent to any of the CHSH inequalities in (\ref{CGLMPset}). Furthermore, it seems likely that this conclusion also applies to any one of the 36 CH-type inequalities given in Ref.~\cite{KKCZO}.

We end this section by noting that the requirement of causal communication does not by itself prevent the sum of probabilities in $I_3$ [cf.~(\ref{I3p})] from reaching its maximum value, $I_3 =4$. Indeed, there exist probability distributions $\{p_1,p_2,\ldots,p_{36}\}$ satisfying \textit{all\/} the constraints in (\ref{NSC1})--(\ref{NC}), and which give $I_3 =4$. An example of such a distribution is:
\begin{equation*}
p_1=p_5=p_9=p_{10}=p_{14}=p_{18}=p_{20}=p_{24}=p_{25}=p_{28}=p_{32}=p_{36}=\frac{1}{3},
\end{equation*}
with all other probabilities zero. This example is the generalization to two three-level systems of the finding by Popescu and Rohrlich \cite{PR} that, for $d=2$, relativistic causality does not constrain the maximum CHSH sum of correlations to $2\sqrt{2}$, but instead it allows for probability distributions giving the maximum level of violation. Arguably, this conclusion generalizes to any dimension $d$.

\section{Optimal set of measurements}

Let us consider a Bell experiment for which the source produces pairs of qutrits in the entangled state:
\begin{equation}
|\psi\rangle = \cos\theta |2\rangle_A |2\rangle_B + \frac{1}{\sqrt{2}}\sin\theta \, 
(|1\rangle_A |1\rangle_B + |3\rangle_A |3\rangle_B ),
\label{state}
\end{equation}
where $\{|1\rangle_{A(B)}, |2\rangle_{A(B)}, |3\rangle_{A(B)}\}$ denotes an orthonormal basis in the state space of qutrit $A$ ($B$). The maximally entangled state is obtained for $\cos\theta=1/\sqrt{3}$ and $\sin\theta=\sqrt{2/3}$. We now describe the set of measurements giving the maximal quantum violation of both the CH and CHSH inequalities \cite{CGLMP,KKCZO,ADGL} (see also Refs.\ \cite{KGZMZ,DKZ,CKKZO}). Firstly, for each of the emitted pairs of qutrits, Alice (Bob) applies a unitary operation $U_{A}^{a}$ ($U_{B}^{b}$), $a,b=1,2$, on qutrit $A$ ($B$), with $U_{A}^{a}$ and $U_{B}^{b}$ given by
\begin{equation}
U_{A}^{a} = \frac{1}{\sqrt{3}}\left[ \begin{array}{ccc}
1 & e^{i\alpha_a} & e^{2i\alpha_a}  \\
1 & \lambda e^{i\alpha_a} & \mu e^{2i\alpha_a}  \\
1 & \mu e^{i\alpha_a} & \lambda e^{2i\alpha_a}
\end{array}  \right] , \quad
U_{B}^{b} = \frac{1}{\sqrt{3}}\left[ \begin{array}{ccc}
1 & e^{i\beta_b} & e^{2i\beta_b}  \\
1 & \mu e^{i\beta_b} & \lambda e^{2i\beta_b}  \\
1 & \lambda e^{i\beta_b} & \mu e^{2i\beta_b}
\end{array}  \right] ,
\label{operations}
\end{equation}
where $\lambda=\exp(2\pi i/3)$, $\mu=\lambda^{\ast}=\exp(4\pi i/3)$, and where $\alpha_a$ ($\beta_b$) is the phase defining $U_{A}^{a}$ ($U_{B}^{b}$). For each run of the experiment, Alice (Bob) has the freedom to choose the transformation $U_{A}^{1}$ or $U_{A}^{2}$ ($U_{B}^{1}$ or $U_{B}^{2}$) to be applied on qutrit $A$ ($B$). The unitary operations in (\ref{operations}) can be realized by means of an unbiased six-port beam splitter \cite{KKCZO}. A detailed description of such devices can be found in Ref.\ \cite{ZZH}. Finally, once $U_{A}^{a}$ and $U_{B}^{b}$ have been applied on the respective qutrit, Alice (Bob) measures the state of the transformed qutrit $A$ ($B$) in the initial basis $\{|1\rangle_{A(B)}, |2\rangle_{A(B)}, |3\rangle_{A(B)}\}$. Thus the joint probability distribution of outcomes predicted by quantum mechanics for the initial state (\ref{state}) is the following:
\pagebreak
\begin{align}
P^{ab}_{\psi}(1,1)=P^{ab}_{\psi}(2,2)=P^{ab}_{\psi}(3,3)=\frac{1}{9}\Big[1&+\sin^2\theta\cos2\phi_{ab}
+\sqrt{2}\sin2\theta\cos\phi_{ab}\Big] ,  \nonumber  \\
P^{ab}_{\psi}(1,2)=P^{ab}_{\psi}(2,3)=P^{ab}_{\psi}(3,1)=\frac{1}{9}\Big[1&-\frac{1}{2}\sin^2\theta\, \big(\!\cos2\phi_{ab}
+\sqrt{3}\sin2\phi_{ab}\big)  \nonumber  \\
&- \frac{1}{\sqrt{2}}\sin2\theta \, \big(\!\cos\phi_{ab}-\sqrt{3}\sin\phi_{ab}\big)\Big] , \label{dist} \\
P^{ab}_{\psi}(1,3)=P^{ab}_{\psi}(2,1)=P^{ab}_{\psi}(3,2)=\frac{1}{9}\Big[1&-\frac{1}{2}\sin^2\theta \, \big(\!\cos2\phi_{ab}
-\sqrt{3}\sin2\phi_{ab}\big)  \nonumber  \\
&- \frac{1}{\sqrt{2}}\sin2\theta \, \big(\!\cos\phi_{ab}+\sqrt{3}\sin\phi_{ab}\big)\Big] ,  \nonumber
\end{align}
where $\phi_{ab}=\alpha_a +\beta_b$. Using the probabilities (\ref{dist}) in (\ref{I3p}), we obtain
\begin{align}
I_{3}(|\psi\rangle) = \frac{\sqrt{3}}{6}\sin^2\theta\, \big( & \sqrt{3} \cos2\phi_{11}+\sin2\phi_{11}
+\sqrt{3}\cos2\phi_{12}-\sin2\phi_{12} \nonumber  \\
& - \sqrt{3} \cos2\phi_{21}- \sin2\phi_{21} +\sqrt{3}\cos2\phi_{22}+ \sin2\phi_{22} \big) \nonumber  \\
+ \, \frac{1}{\sqrt{6}}& \sin  2\theta \, \big( \sqrt{3} \cos\phi_{11}-\sin\phi_{11}
+\sqrt{3}\cos\phi_{12}+\sin\phi_{12} \nonumber  \\
& - \sqrt{3} \cos\phi_{21}+ \sin\phi_{21} +\sqrt{3}\cos\phi_{22}- \sin\phi_{22} \big).
\label{I3pp}
\end{align}
Of course, from the results of the preceding section, the quantum prediction for $K_3$ will be given by $K_{3}(|\psi\rangle)=(I_{3}(|\psi\rangle)-2)/3$, as one can check directly by using the joint probabilities (\ref{dist}) in either (\ref{K3}) or (\ref{K3p}), and putting the single probabilities equal to $\frac{1}{3}$.

Consider now the following values of the phases
\begin{equation}
\alpha_2 = \alpha_1 +\frac{\pi}{3}, \qquad \beta_1 = -\alpha_1 +\frac{\pi}{6},
\qquad \beta_2 = -\alpha_1 -\frac{\pi}{6},
\label{settings}
\end{equation}
where $\alpha_2$, $\beta_1$, and $\beta_2$ are given in terms of the variable phase $\alpha_1$. For the settings in (\ref{settings}), expression (\ref{I3pp}) reduces to
\begin{equation}
I_{3}(|\psi\rangle) = 2\sin^2 \theta + 2\sqrt{\frac{2}{3}} \sin 2\theta ,
\label{I3theta}
\end{equation}
which is independent of $\alpha_1$. The corresponding expression for $K_3$ is
\begin{equation}
K_{3}(|\psi\rangle) = \frac{2}{3} \Big( \sqrt{\frac{2}{3}} \sin 2\theta -\cos^2 \theta \Big).
\label{K3theta}
\end{equation}
In Fig.\ 1, the functions in (\ref{I3theta}) and (\ref{K3theta}) have been plotted for $0\leq \theta \leq \pi$. The maximum values are
\begin{equation*}
I_{3}^{\text{max}}(|\psi\rangle)= 1+\sqrt{11/3}\approx 2.915,
\end{equation*}
and 
\begin{equation*}
K_{3}^{\text{max}}(|\psi\rangle)=(\sqrt{11/3}-1)/3 \approx 0.305,
\end{equation*}
and they are attained for an angle $\theta_{\text{max}}\approx 60.74^{\circ}$. Explicitly, the state leading to the maximal violation is
\begin{equation}
|\psi_{\text{mv}}\rangle = \sqrt{\frac{11-\sqrt{33}}{22}}\,\, |2\rangle_A |2\rangle_B 
+ \sqrt{\frac{11+\sqrt{33}}{44}} \,\,(|1\rangle_A |1\rangle_B + |3\rangle_A |3\rangle_B ).
\label{mv} 
\end{equation}
Previous numerical work \cite{ADGL} shows the optimality of the chosen set of measurements, the values $I_{3}^{\text{max}}(|\psi\rangle)$ and $K_{3}^{\text{max}}(|\psi\rangle)$ indeed being the maximum ones predicted by quantum mechanics for the case in which two von Neumann measurements are performed by each of the parties. In particular, these values are slightly larger than those obtained for the maximally entangled state, namely,
\pagebreak
\begin{align*}
I_{3}(|\psi_{\text{me}}\rangle) &= (12+8\sqrt{3})/9 \approx 2.873,  \\ K_{3}(|\psi_{\text{me}}\rangle) &= (8\sqrt{3}-6)/27 \approx 0.291. 
\end{align*}
Note that the inequalities are not violated by the states in (\ref{state}) for which $\theta =n\pi /2$ ($n=0,\pm1, \pm2,\ldots\,$).
\begin{figure}[ttt]
\vspace{-1cm}
\centering
\captionstyle{centerlast}
\hspace{-.5cm}
\includegraphics[width=3.7in]{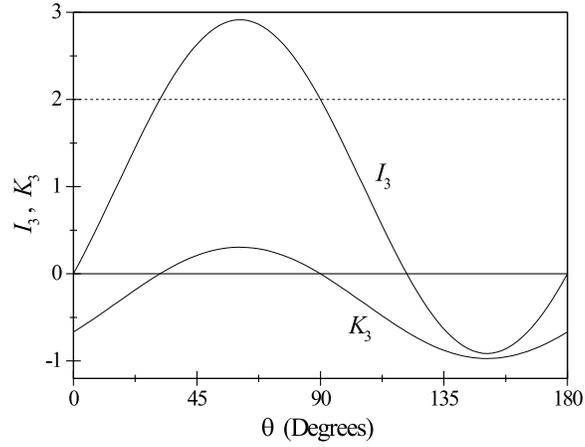}
\setlength{\abovecaptionskip}{-.3cm}
\setlength{\textfloatsep}{9pt plus 2pt minus 3pt}
\setcaptionmargin{1.5cm}
\renewcommand{\figurename}{Fig.}
\renewcommand{\captionlabeldelim}{.~}
\caption{\small{$I_3(\theta)$ and $K_3(\theta)$ as predicted by quantum mechanics for pairs of qutrits in the state (\ref{state}), where $I_3(\theta)$ and $K_3(\theta)$ are evaluated in the case of condition (\ref{settings}). Such functions are related to each other by $I_3 = 2+ 3K_3$.}}
\centering
\end{figure} When $n$ is even such states correspond to product states, and when $n$ is odd they correspond to maximally entangled states of the form $(1/\sqrt{2})(|1\rangle_A |1\rangle_B + |3\rangle_A |3\rangle_B )$. The latter state describes two entangled qutrits each of them living in a two-dimensional state space. Such a state does not exploit the full dimensionality of the qutrits space, and hence it cannot violate the inequalities. On the other hand, the states in (\ref{state}) that, for the measurements considered, yield a violation of either the CHSH inequality ($I_3 \leq 2$) or the CH inequality ($K_3 \leq 0$) are those for which $\arctan\sqrt{3/8} < \theta < \pi/2$ (mod $\pi$).

If the initial state (\ref{state}) is mixed with some amount of uncolored noise, the state becomes
\begin{equation}
\rho = \lambda \, |\psi\rangle \langle\psi| + (1-\lambda)\, \frac{\openone}{9},
\end{equation}
where $0\leq \lambda \leq 1$. Quantum mechanics now predicts the probabilities:
\begin{align*}
P^{ij}_{\rho}(a_i,b_j) &= \lambda P^{ij}_{\psi}(a_i,b_j)+\frac{1-\lambda}{9}, \\
P^{i}_{\rho}(a_i) &= \lambda P^{i}_{\psi}(a_i)+\frac{1-\lambda}{3}, \\
Q^{j}_{\rho}(b_j) &= \lambda Q^{j}_{\psi}(b_j)+\frac{1-\lambda}{3}.
\end{align*}
Since $P^{i}_{\psi}(a_i) = Q^{j}_{\psi}(b_j) = \tfrac{1}{3}$, then the same holds true for the new single probabilities, $P^{i}_{\rho}(a_i) = Q^{j}_{\rho}(b_j) = \tfrac{1}{3}$. Correspondingly, $I_3$ and $K_3$ change to $I_3(\rho) = \lambda I_3(|\psi\rangle)$ and $K_3(\rho) = \lambda K_3(|\psi\rangle)-\tfrac{2(1-\lambda)}{3}$, respectively. Therefore, for the case in which $I_3(|\psi\rangle)>2$, the inequality $I_3\leq 2$ will be violated by quantum mechanics if and only if
\begin{equation}
\lambda > \frac{2}{I_3(|\psi\rangle)},
\label{lambda1}
\end{equation}
and, similarly, for the case in which $K_3(|\psi\rangle)>0$, the inequality $K_3\leq 0$ will be violated by quantum mechanics if and only if
\begin{equation}
\lambda > \frac{2}{2+3K_3(|\psi\rangle)}.
\label{lambda2}
\end{equation}
Note that, as expected, the conditions in (\ref{lambda1}) and (\ref{lambda2}) are exactly the same since $I_3(|\psi\rangle)=2+3K_3(|\psi\rangle)$. So there exists a critical value $\lambda=2/I_3(|\psi\rangle)$ above which a local realistic description of the experiment is not possible. The optimal, minimum value of $\lambda$ is obtained when $I_3(|\psi\rangle)$ is maximum, i.e.
\begin{equation*}
\lambda_{\text{min}}=2/I_{3}^{\text{max}}(|\psi\rangle) = (\sqrt{33}-3)/4 \approx 0.686,
\end{equation*}
and this optimal value being achieved for the state $|\psi_{\text{mv}}\rangle$. Put it another way, the maximum amount of uncolored noise that can be added to the two-qutrit system while still getting a violation of Bell's inequality is:
\begin{equation*}
1-\lambda_{\text{min}}=(7-\sqrt{33})/4 \approx 0.314.
\end{equation*}

We conclude this section by noting that one could equally measure the strength of the inequality $I_3 \leq 2$ or $K_3 \leq 0$ by mixing the initial state with some kind of noise other than uncolored noise. For example, one could consider the possibility of mixing the initial entangled state with the closest separable one, or to mix it with the tensor product state of the reduced density matrices. Remarkably, it turns out \cite{ADGL} that the optimal values of $\lambda$, $\lambda_{\text{min}}^{\prime}$ and $\lambda_{\text{min}}^{\prime\prime}$, provided by these alternative measures of nonlocality for the state $|\psi_{\text{mv}}\rangle$ coincide, and they are equal to the optimal value obtained when $|\psi_{\text{mv}}\rangle$ is mixed with some amount of (uncolored) noise, i.e. $\lambda_{\text{min}}=\lambda_{\text{min}}^{\prime}=\lambda_{\text{min}}^{\prime\prime}$.

\section{Finite detector efficiency}

Now we consider the case in which each of the detectors in our Bell experiment is endowed with a quantum efficiency $\eta$ ($0\leq \eta \leq 1$), where $\eta$ is meant to be the probability that a detector ``cliks'' when a particle (qutrit) impinges on it. (We note that, quite generally, $\eta$ may also account for all possible losses of the particles on their way from the source to the detectors.) Quantum mechanics then predicts the modified probabilities:
\begin{align*}
P^{ij}_{\eta}(a_i,b_j) &= \eta^2 P^{ij}_{\psi}(a_i,b_j), \\
P^{i}_{\eta}(a_i) &= \eta P^{i}_{\psi}(a_i), \\
Q^{j}_{\eta}(b_j) &= \eta Q^{j}_{\psi}(b_j),
\end{align*}
so the new single probabilities are given by $P^{i}_{\eta}(a_i) =Q^{j}_{\eta}(b_j) =\tfrac{\eta}{3}$. Correspondingly, the quantum prediction for $I_3$ and $K_3$ becomes
\begin{gather}
I^{\eta}_3(|\psi\rangle) = \eta^2 I_3(|\psi\rangle), \label{etaI3} \\
K^{\eta}_3(|\psi\rangle) = \eta^2 K_3(|\psi\rangle) +\frac{4}{3} \eta(\eta -1),
\label{etaK3}
\end{gather}
where $I_3(|\psi\rangle)$ and $K_3(|\psi\rangle)$ represent the quantum prediction of $I_3$ and $K_3$ evaluated for $\eta =1$. Thus, for the case in which $I_3(|\psi\rangle)>2$, it follows from (\ref{etaI3}) that the inequality $I_3 \leq 2$ will be violated by quantum mechanics if and only if
\begin{equation}
\eta > \sqrt{\frac{2}{I_3(|\psi\rangle)}}.
\label{eta1}
\end{equation}
\begin{figure}[ttt]
\vspace{-.5cm}
\centering
\captionstyle{centerlast}
\hspace{-.2cm}
\includegraphics[width=3.3in]{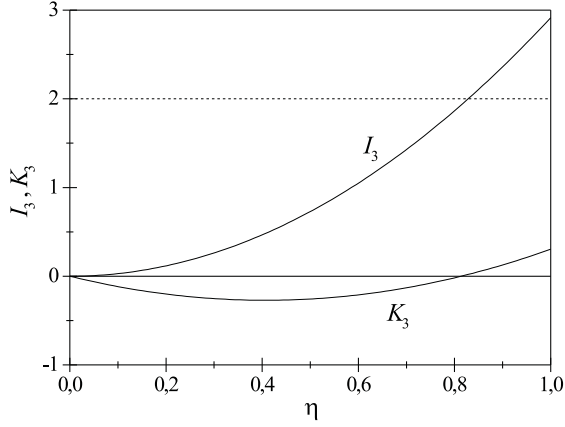}
\setlength{\abovecaptionskip}{-.3cm}
\setlength{\textfloatsep}{9pt plus 2pt minus 3pt}
\setcaptionmargin{1.5cm}
\renewcommand{\figurename}{Fig.}
\renewcommand{\captionlabeldelim}{.~}
\caption{\small{Quantum prediction of $I_3$ and $K_3$ for the state $|\psi_{\text{mv}}\rangle$ and the optimal settings (\ref{settings}), as a function of the efficiency parameter $\eta$. The inequality $I_3 \leq 2$ ($K_3 \leq 0$) is violated provided $\eta$ exceeds the threshold value $\eta^{\text{CHSH}}_{0}\approx 0.828$ ($\eta^{\text{CH}}_{0}\approx 0.814$).}}
\centering
\end{figure}Similarly, for the case in which $K_3(|\psi\rangle)>0$, it follows from (\ref{etaK3}) that the inequality $K_3 \leq 0$ will be violated by quantum mechanics if and only if
\begin{equation}
\eta > \frac{4}{4+3K_3(|\psi\rangle)}=\frac{4}{2+I_3(|\psi\rangle)}.
\label{eta2}
\end{equation}
From (\ref{eta1}) and (\ref{eta2}), we can see that the condition for the violation of the inequality $I_3 \leq 2$ differs from the condition for the violation of the inequality $K_3 \leq 0$. This is a consequence of the fact that, actually, the inequalities $I_3 \leq 2$ and $K_3 \leq 0$ themselves are not equivalent for $\eta <1$. Indeed, from (\ref{etaI3}) and (\ref{etaK3}), and the relation $I_{3}(|\psi\rangle)= 2+3K_3(|\psi\rangle)$, it follows that
\begin{equation}
I^{\eta}_3(|\psi\rangle) = 3K^{\eta}_3(|\psi\rangle) +4\eta -2\eta^2.
\label{I3K3}
\end{equation}
So, from (\ref{I3K3}), we can see that $I^{\eta}_3(|\psi\rangle) \neq 2+3K^{\eta}_3(|\psi\rangle)$ as soon as $\eta <1$. Therefore, it can be deduced from this that the inequalities $I_3 \leq 2$ and $K_3 \leq 0$ are no longer equivalent whenever $\eta <1$. Note that, in any case, the no-signaling condition is satisfied by the modified probabilities $P^{ij}_{\eta}(a_i,b_j) = \eta^2 P^{ij}_{\psi}(a_i,b_j)$, as the constraints in (\ref{NSC1})--(\ref{NSC2}) are already fulfilled by the $P^{ij}_{\psi}(a_i,b_j)$'s.

In Fig.\ 2, $I^{\eta}_3(|\psi\rangle)$ and $K^{\eta}_3(|\psi\rangle)$ have been plotted as a function of $\eta$ for the case in which $|\psi\rangle$ is the state in (\ref{mv}). As we saw in the preceding section, for this state quantum mechanics predicts the maximum value of $I_3$ to be $I_3(|\psi_{\text{mv}}\rangle)=1+\sqrt{11/3}$. Substituting this value into (\ref{eta1}) and (\ref{eta2}), we find that the minimum threshold value of $\eta$, allowing for a violation of the CHSH inequality $I_3 \leq 2$, to be $\eta^{\text{CHSH}}_{0}=(\sqrt{-3+\sqrt{33}})/2 \approx 0.828$, while that allowing for a violation of the CH inequality, $K_3 \leq 0$, is $\eta^{\text{CH}}_{0}=(9-\sqrt{33})/4 \approx 0.814$. For an arbitrary state $|\psi\rangle$, the threshold values acquire the general form (see (\ref{eta1}) and (\ref{eta2})),
\begin{gather}
\eta^{\text{CHSH}}= \sqrt{\frac{2}{I_3(|\psi\rangle)}}, \label{eta3}  \\
\eta^{\text{CH}}= \frac{4}{2+I_3(|\psi\rangle)}.  \label{eta4}
\end{gather}
In Fig.\ 3, $\eta^{\text{CHSH}}$ and $\eta^{\text{CH}}$ have been plotted for the case in which $I_3(|\psi\rangle)$ is given by (\ref{I3theta}), where now $\theta$ is restricted to vary within the interval $\arctan\sqrt{3/8} < \theta < \pi/2$. (These values of $\theta$ correspond to the states in (\ref{state}) yielding a violation of either the CHSH or CH inequalities for the case of ideal detectors; see Fig.~1.) Note that $\eta^{\text{CHSH}} > \eta^{\text{CH}}$ for the entire range of variation, the maximum difference between the values of $\eta^{\text{CHSH}}$ and $\eta^{\text{CH}}$ occurring for the state $|\psi_{\text{mv}}\rangle$. So, for a detector efficiency $\eta$ lying in the interval $\eta^{\text{CH}}_{0} < \eta < \eta^{\text{CHSH}}_{0}$, and for an ensemble of pairs of qutrits in the state $|\psi_{\text{mv}}\rangle$, it would be possible to rule out local realism by using the CH inequality, but such a refutation would not be possible if one instead uses the CHSH inequality. Please notice that, in any case, the difference $\eta^{\text{CHSH}}-\eta^{\text{CH}}$ is practically negligible, and that even the optimal values $\eta^{\text{CH}}_{0}$ and $\eta^{\text{CHSH}}_{0}$ are quite demanding to be currently achieved in practice for the usual case of detectors registering optical photons. Experiments with two six-port beam splitters of the type described in Refs.~\cite{KKCZO} and \cite{ZZH} aimed to test either the inequality $I_3 \leq 2$ or $K_3 \leq 0$, must therefore rely on one or another sort of supplementary assumption (like the fair-sampling assumption) in order to deal with the problem of the low detection efficiencies.
\begin{figure}[ttt]
\vspace{-.4cm}
\centering
\captionstyle{centerlast}
\hspace{-.4cm}
\includegraphics[width=3.6in]{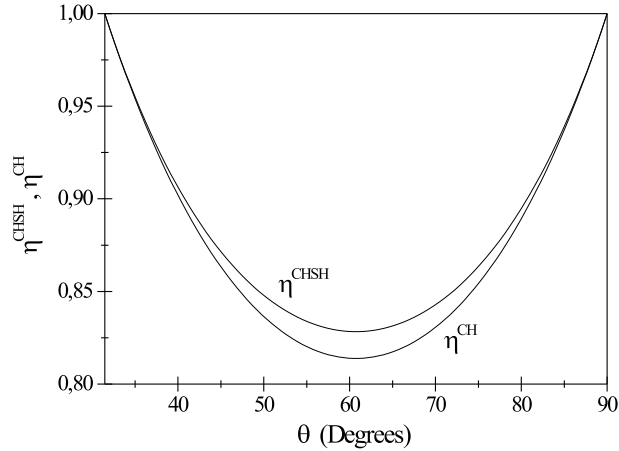}
\setlength{\abovecaptionskip}{-.3cm}
\setlength{\textfloatsep}{9pt plus 2pt minus 3pt}
\setcaptionmargin{1.5cm}
\renewcommand{\figurename}{Fig.}
\renewcommand{\captionlabeldelim}{.~}
\caption{\small{$\eta^{\text{CHSH}}$ ($\eta^{\text{CH}}$): Threshold values of $\eta$ allowing for a violation of the inequality $I_3 \leq 2$ ($K_3 \leq 0$) for the subset of states in Eq.\ (\ref{state}) for which $\arctan\sqrt{3/8} < \theta < \pi/2$. The curves are symmetric with respect to the axis $\theta_0\approx 60.74^{\circ}$ corresponding to the state $|\psi_{\text{mv}}\rangle$.}}
\centering
\end{figure}

We should add here that the threshold value in (\ref{eta4}) was already derived in an independent way by Massar \textit{et al.} \cite{MPRG}. Indeed, in order to reproduce analytically the optimal numerical values of $\eta$ obtained in Ref.\ \cite{MPRG} (which are identical to those given in Ref.\ \cite{DKZ}) for the case of systems of arbitrary dimension and two settings (measurements) on each side, Massar \textit{et al.} derived the following Bell inequality:
\begin{equation}
S_d \equiv I_d + \frac{1}{2} \sum_{i,j=1}^{2} P(A_i=\emptyset,B_j=\emptyset) \leq 2,
\label{Sd}
\end{equation}
where $I_d$ is the Bell expression for $d$-dimensional systems given in Ref.\ \cite{CGLMP}, and $P(A_i=\emptyset,B_j=\emptyset)$ denotes the probability of a double non-detection event, that is, the probability that the detectors fail to register each of the two entangled particles emitted by the source. (We are assuming here that the probability $\lambda$ that the pair of particles is produced by the source of entangled systems is unity, although, as shown in Ref.\ \cite{MPRG}, the threshold detection efficiency remains unchanged for the case in which $\lambda <1$, as far as inequality (\ref{Sd}) is concerned.) Now, replacing $I_d$ and $P(A_i=\emptyset,B_j=\emptyset)$ by the quantum predictions $\eta^2 I_d(|\psi\rangle)$ and $(1-\eta)^2$, respectively, one easily finds that the condition for violation by quantum mechanics of the Bell inequality (\ref{Sd}) is that $\eta$ must be greater than
\begin{equation}
\eta = \frac{4}{2+I_d(|\psi\rangle)},
\label{eta5}
\end{equation}
which, for $d=3$, reduces to (\ref{eta4}).

To end this section we show that, for the considered case $d=3$, if quantum mechanics predicts that $S_3(|\psi\rangle)>2$, then necessarily $K_3(|\psi\rangle)>0$, and vice versa. (The previous statement is of course tantamount to saying that, if $S_3(|\psi\rangle)\leq 2$, then necessarily $K_3(|\psi\rangle)\leq 0$, and vice versa.) This follows quickly by simply noting that, when substituting the quantum predictions $\eta^2 I_3(|\psi\rangle)$ and $(1-\eta)^2$ for $I_3$ and $P(A_i=\emptyset,B_j=\emptyset)$, respectively, inequality (\ref{Sd}) can be written in the form
\begin{equation}
\eta^2 I_3(|\psi\rangle) + 2\eta^2 -4\eta \leq 0.
\label{Sd1}
\end{equation}
Now, from (\ref{I3K3}), we have that
\begin{equation*}
I^{\eta}_3(|\psi\rangle) = \eta^2 I_3(|\psi\rangle)= 3K^{\eta}_3(|\psi\rangle)+4\eta -2\eta^2,
\end{equation*}
and then inequality (\ref{Sd1}) reduces to $3K^{\eta}_3(|\psi\rangle)\leq 0$, or $K^{\eta}_3(|\psi\rangle)\leq 0$. Therefore, in view of (\ref{eta4}) and (\ref{eta5}), and from the numerical results obtained in Ref.\ \cite{MPRG}, it seems safe to conclude that the CH inequality $K_3\leq 0$ is optimal with respect to inefficient detectors.

\section{Conclusions}

We have argued that each of the CHSH inequalities $I_3\leq 2$ in (\ref{CGLMPset}) is associated with one independent CH inequality $K_3 \leq 0$ through the relation $I_3 = 2+3K_3$. Clearly, if $I_3 = 2+3K_3$, then the CH inequality $K_3\leq 0$ implies the CHSH inequality $I_3 \leq 2$, and conversely, the CHSH inequality $I_3 \leq 2$ implies the CH inequality $K_3\leq 0$. Such correspondence has been shown explicitly for the  pairs of inequalities $I_3\leq2$ and $K_3\leq0$ [cf.~(\ref{I3}) and (\ref{K3})], and  $I_3^{\prime}\leq2$ and $K_3^{\prime}\leq0$ [cf.~(\ref{I3prime}) and (\ref{K3prime})]. A similar relationship could be established between each CHSH inequality in the set (\ref{CGLMPset}) and the corresponding CH inequality. It thus follows that, for ideal detectors, if the CHSH inequality $I_3 \leq 2$ turns out to be the necessary and sufficient condition for the existence of a local and realistic model reproducing the predicted probabilities, then the same is true for the corresponding CH inequality $K_3 \leq 0$, and vice versa. We have argued, on the other hand, that the set of CH inequalities introduced in Ref.\ \cite{KKCZO} is not equivalent to any of the CHSH inequalities (\ref{CGLMPset}). Specifically, we have shown that the inequality $W_3 \leq 0$ in (\ref{W3}) and the inequality $I_3 \leq 2$ in (\ref{I3}) do not mutually imply each other.

We have seen that, as first shown in Ref.~\cite{ADGL}, the maximum quantum violation of either the inequality $I_3\leq 2$ or $K_3\leq 0$ for the case of two von Neumann measurements per site, is obtained for the nonmaximally entangled state $|\psi_{\text{mv}}\rangle$, (\ref{mv}). Furthermore, both inequalities $I_3\leq 2$ and $K_3\leq 0$ exhibit the same resistance to the addition of uncolored noise [cf.~(\ref{lambda1}) and (\ref{lambda2})]. The equivalence between such inequalities, however, is lost when the less-than-perfect detector efficiency is taken into account. Indeed, we have seen that $I_3 \neq 2+3K_3$ for $\eta <1$. For the optimal set of measurements, we have calculated the minimum detector efficiency $\eta^{\text{CH}}_{0}$ ($\eta^{\text{CHSH}}_{0}$) necessary for quantum mechanics to violate the CH (CHSH) inequalities. The fact that $\eta^{\text{CH}}_{0} \neq \eta^{\text{CHSH}}_{0}$ indicates that the inequalities $K_3 \leq 0$ and $I_3 \leq 2$ are no longer equivalent whenever $\eta < 1$. In this respect, we should mention that the fact that the CHSH inequalities giving overestimated values of threshold quantum efficiencies with respect to the CH inequalities does not allow us to conclude that, ``$\ldots$these inequalities [the CHSH inequalities] are only a necessary condition for local realism'' \cite{KKCZO}. This is so because the conditions $I_3 \leq 2$ and $K_3 \leq0$ entailed by such inequalities apply to two \textit{different\/} experiments---one testing the CHSH inequality, and the other one testing the CH inequality---so that such conditions can be considered to be essentially independent of each other.

\vspace{1cm}

\noindent
\textbf{\large{Appendix A.}}
\vspace{.3cm}

\noindent
Here we give the probabilities $p_2$, $p_9$, $p_{13}$, $p_{18}$, $p_{20}$, $p_{24}$, $p_{25}$, $p_{28}$, $p_{32}$, and $p_{36}$ obtained by solving the system of equations (\ref{NSC1})--(\ref{NC}) with respect to the variables $\{p_2,p_4,p_9,p_{11},p_{13},p_{18},p_{20},p_{24},p_{25},p_{28},p_{32},p_{36}\}$. These are
\begin{align}
p_{2} = \,\, & \frac{1}{3} \big( 1-2p_1-p_3-p_5+p_6-p_7-2p_8+2p_{10}+p_{12}
      -2p_{14}-p_{15}+p_{16}-p_{17} \nonumber \\ 
    & -p_{19}-2p_{21}+p_{22}+2p_{23}+p_{26}-p_{27}
      +2p_{29}+p_{30}-2p_{31}-p_{33}-p_{34}+p_{35} \big), \tag{A.1} \\[.1cm]
p_{9} = \,\, & \frac{1}{3} \big( 1+p_1-p_3-p_5-2p_6-p_7-2p_8-p_{10}-2p_{12}
      +p_{14}-p_{15}+p_{16}+2p_{17} \nonumber \\ 
    & -p_{19}+p_{21}-2p_{22}-p_{23}+p_{26}+2p_{27}
      -p_{29}+p_{30}+p_{31}+2p_{33}-p_{34}-2p_{35} \big), \tag{A.2} \\[.1cm]
p_{13} = \,\, & \frac{1}{3} \big( 1-2p_1-p_3+2p_5+p_6-p_7+p_8-p_{10}+p_{12}
      -2p_{14}-p_{15}-2p_{16}-p_{17} \nonumber \\ 
    & +2p_{19}+p_{21}+p_{22}-p_{23}-2p_{26}-p_{27}
      -p_{29}-2p_{30}+p_{31}-p_{33}+2p_{34}+p_{35} \big), \tag{A.3} \\[.1cm]
p_{18} = \,\, & \frac{1}{3} \big( 1+p_1-p_3-p_5-2p_6+2p_7+p_8-p_{10}-2p_{12}
      +p_{14}-p_{15}-2p_{16}-p_{17} \nonumber \\ 
    & -p_{19}+p_{21}-2p_{22}-p_{23}+p_{26}+2p_{27}
      -p_{29}+p_{30}+p_{31}+2p_{33}-p_{34}-2p_{35} \big),  \tag{A.4}
\end{align}
\begin{align}
p_{20} = \,\, & \frac{1}{3} \big( 1-2p_1-p_3+2p_5+p_6-p_7+p_8+2p_{10}+p_{12}
      -2p_{14}-p_{15}+p_{16}-p_{17} \nonumber \\ 
    & -p_{19}-2p_{21}+p_{22}-p_{23}-2p_{26}-p_{27}
      +2p_{29}+p_{30}-2p_{31}-p_{33}-p_{34}+p_{35} \big), \tag{A.5}   \\[.1cm]
p_{24} = \,\, & \frac{1}{3} \big( 1+p_1+2p_3-p_5+p_6-p_7-2p_8-p_{10}-2p_{12}
      +p_{14}-p_{15}+p_{16}+2p_{17} \nonumber \\ 
    & -p_{19}-2p_{21}-2p_{22}-p_{23}+p_{26}-p_{27}
      -p_{29}+p_{30}+p_{31}+2p_{33}-p_{34}-2p_{35} \big), \tag{A.6}  \\[.1cm]
p_{25} = \,\, & \frac{1}{3} \big( 1+p_1-p_3-p_5-2p_6+2p_7+p_8-p_{10}+p_{12}
      +p_{14}+2p_{15}-2p_{16}-p_{17} \nonumber \\ 
    & -p_{19}+p_{21}-2p_{22}-p_{23}-2p_{26}-p_{27}
      -p_{29}-2p_{30}+p_{31}-p_{33}+2p_{34}+p_{35} \big), \tag{A.7} \\[.1cm]
p_{28} = \,\, & \frac{1}{3} \big( 1-2p_1-p_3+2p_5+p_6-p_7+p_8+2p_{10}+p_{12}
      -2p_{14}-p_{15}+p_{16}-p_{17} \nonumber \\ 
    & +2p_{19}+p_{21}+p_{22}-p_{23}-2p_{26}-p_{27}
      -p_{29}-2p_{30}-2p_{31}-p_{33}-p_{34}+p_{35} \big), \tag{A.8}  \\[.1cm]
p_{32} = \,\, & \frac{1}{3} \big( 1+p_1+2p_3-p_5+p_6-p_7-2p_8-p_{10}-2p_{12}
      +p_{14}-p_{15}+p_{16}+2p_{17} \nonumber \\ 
    & -p_{19}-2p_{21}+p_{22}+2p_{23}+p_{26}-p_{27}
      -p_{29}+p_{30}-2p_{31}-p_{33}-p_{34}-2p_{35} \big), \tag{A.9}  \\[.1cm]
p_{36} = \,\, & \frac{1}{3} \big( 1+p_1-p_3-p_5-2p_6+2p_7+p_8-p_{10}+p_{12}
      +p_{14}+2p_{15}-2p_{16}-p_{17} \nonumber \\ 
    & -p_{19}+p_{21}-2p_{22}-p_{23}+p_{26}+2p_{27}
      -p_{29}-2p_{30}+p_{31}-p_{33}-p_{34}-2p_{35} \big).  \tag{A.10}
\end{align}


\vspace{.1cm}



\begin{thebibliography}{99}


\bibitem{Bell64} J. S. Bell, \textit{Physics} (Long Island City, N.Y.) \textbf{1}, 195 (1964).

\bibitem{CGLMP} D. Collins, N. Gisin, N. Linden, S. Massar, and S. Popescu, \textit{Phys. Rev. Lett.} \textbf{88}, 040404 (2002).

\bibitem{CHSH} J. F. Clauser, M. A. Horne, A. Shimony, and R. A. Holt, \textit{Phys. Rev. Lett.} \textbf{23}, 880 (1969).

\bibitem{KKCZO} D. Kaszlikowski, L. C. Kwek, J.-L. Chen, M. \.{Z}ukowski, and C. H. Oh, \textit{Phys. Rev. A} \textbf{65}, 032118 (2002); e-print quant-ph/0106010.

\bibitem{CH} J. F. Clauser and M. A. Horne, \textit{Phys. Rev. D} \textbf{10}, 526 (1974).

\bibitem{ZZH} M. \.{Z}ukowski, A. Zeilinger, and M. A. Horne, \textit{Phys. Rev. A} \textbf{55}, 2564 (1997).

\bibitem{WZPW} X.-H. Wu, H.-S. Zong, H.-R. Pang, and F. Wang, \textit{Phys. Lett. A} \textbf{281}, 203 (2001).

\bibitem{KGZMZ} D. Kaszlikowski, P. Gnaci\'{n}ski, M. \.{Z}ukowski, W. Miklaszewski, and A. Zeilinger, \textit{Phys. Rev. Lett.} \textbf{85}, 4418 (2000).

\bibitem{DKZ} T. Durt, D. Kaszlikowski, and M. \.{Z}ukowski, \textit{Phys. Rev. A} \textbf{64}, 024101 (2001).

\bibitem{CKKZO} J.-L. Chen, D. Kaszlikowski, L. C. Kwek, C. H. Oh, and M. \.{Z}ukowski, \textit{Phys. Rev. A} \textbf{64}, 052109 (2001).

\bibitem{Mermin} N. D. Mermin, \textit{Fundamental Problems in Quantum Theory\/}, eds. D. M. Greenberger and A.~Zeilinger, \textit{Ann. N.Y. Acad. Sci.\/} \textbf{755}, 616 (1995).

\bibitem{Cereceda} J. L. Cereceda, \textit{Found. Phys. Lett.} \textbf{14}, 401 (2001).

\bibitem{Masanes} L. Masanes, e-print quant-ph/0210073.

\bibitem{PR} S. Popescu and D. Rohrlich, \textit{Found. Phys.} \textbf{24}, 379 (1994).

\bibitem{ADGL} A. Ac\'{i}n, T. Durt, N. Gisin, and J. I. Latorre, \textit{Phys. Rev. A} \textbf{65}, 052325 (2002).

\bibitem{MPRG} S. Massar, S. Pironio, J. Roland, and B. Gisin, \textit{Phys. Rev. A} \textbf{66}, 052112 (2002).


\end{thebibliography}
\end{document}